\documentclass[a4paper,11pt]{article}
\pdfoutput=1 

\usepackage{jcappub} 

\usepackage[T1]{fontenc} 

\makeatletter
\def\@fpheader{\relax}
\makeatother

\title{\boldmath Fate of the \textit{true-vacuum} bubbles}

\author[a]{J. A. Gonz\'alez,}
\author[b]{A. Bellor\'in,}
\author[c]{M\'onica A. Garc\'ia-\~Nustes,}
\author[d]{L. E. Guerrero,}
\author[e]{S. Jim\'enez,}
\author[c,1]{Juan F. Mar\'in,\note{Corresponding author.}}
\author[f]{and L. V\'azquez}

\affiliation[a]{Department of Physics, Florida International University,\\ Miami, Florida 33199, USA}
\affiliation[b]{Escuela de F\'isica, Facultad de Ciencias, Universidad Central de Venezuela,\\ Apartado Postal 47586, Caracas 1041-A, Venezuela}
\affiliation[c]{Instituto de F\'isica, Pontificia Universidad Cat\'olica de Valpara\'iso,\\ Casilla 4059, Chile}
\affiliation[d]{Departamento de F\'isica, Universidad Sim\'on Bol\'ivar,\\ Apartado Postal 89000, Caracas 1080-A, Venezuela}
\affiliation[e]{Departamento de Matem\'atica Aplicada a las TT.II., E.T.S.I. Telecomunicaci\'on,\\ Universidad Polit\'ecnica de Madrid, 28040-Madrid, Spain}
\affiliation[f]{Departamento de Matem\'atica Aplicada, Facultad de Inform\'atica,\\ Universidad Complutense de Madrid, 28040-Madrid, Spain}

\emailAdd{jorgalbert3047@hotmail.com}
\emailAdd{alberto.bellorin@ucv.ve}
\emailAdd{monica.garcia@pucv.cl}
\emailAdd{lguerre@usb.ve}
\emailAdd{s.jimenez@upm.es}
\emailAdd{juanfmarinm@gmail.com}
\emailAdd{lvazquez@fdi.ucm.es}

\abstract{We investigate the bounce solutions in vacuum decay problems. We show that it is possible to have a stable false vacuum in a potential
that is unbounded from below.}

\begin{document}
\maketitle
\flushbottom

\section{Introduction}
\label{intro}

Is our vacuum metastable? Is the Universe that we live in inherently unstable?

\begin{figure}[tbp]
\centering
\includegraphics[width=.49\textwidth]{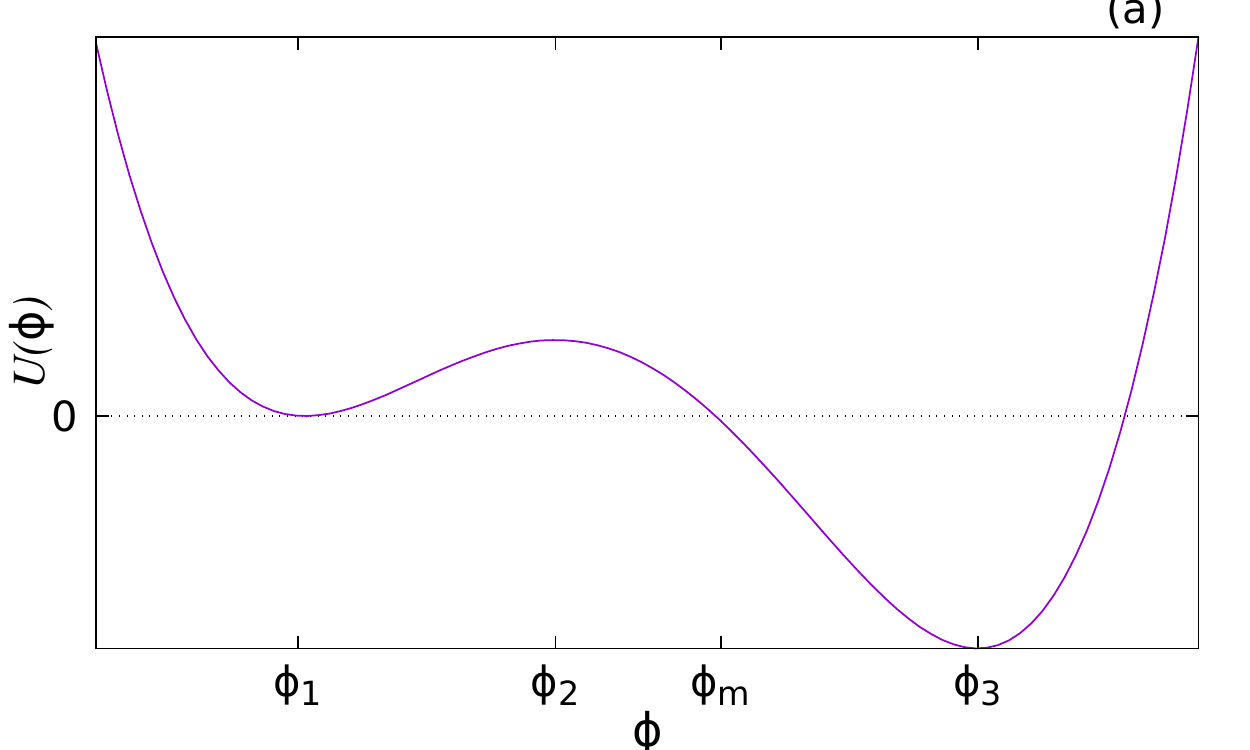}
\hfill
\includegraphics[width=.49\textwidth]{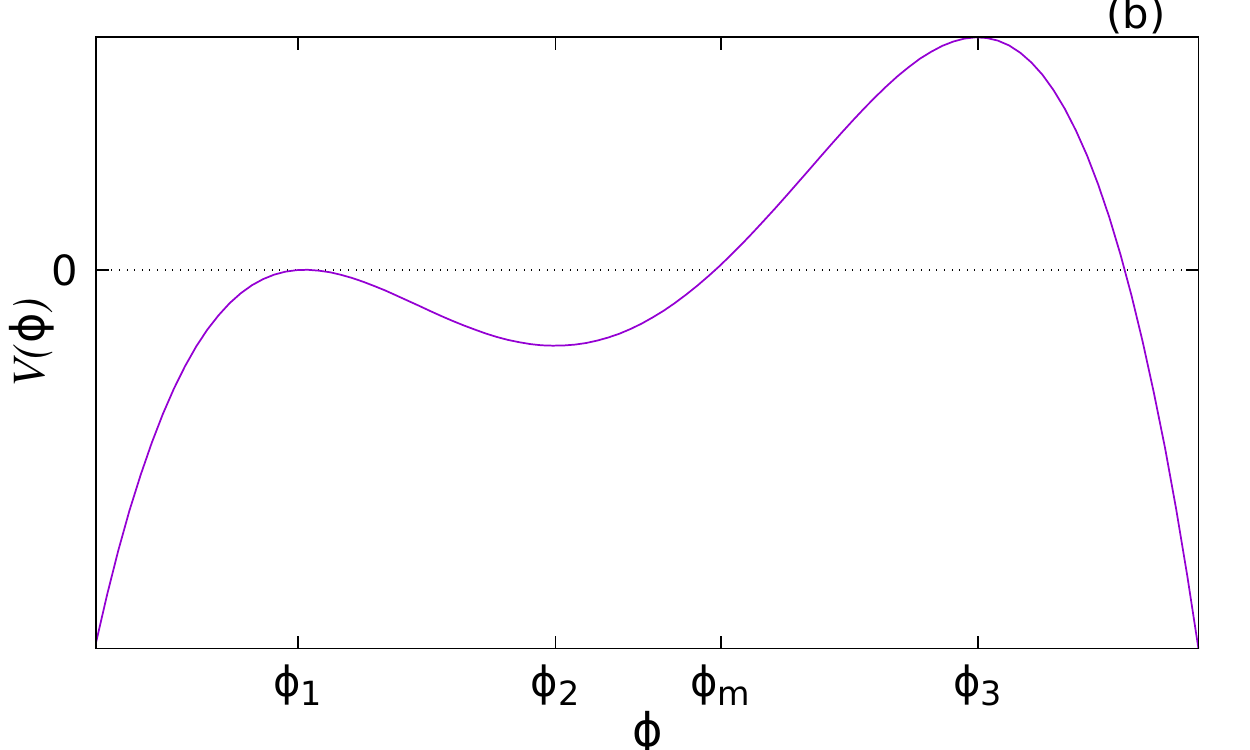}\\
\includegraphics[width=.49\textwidth]{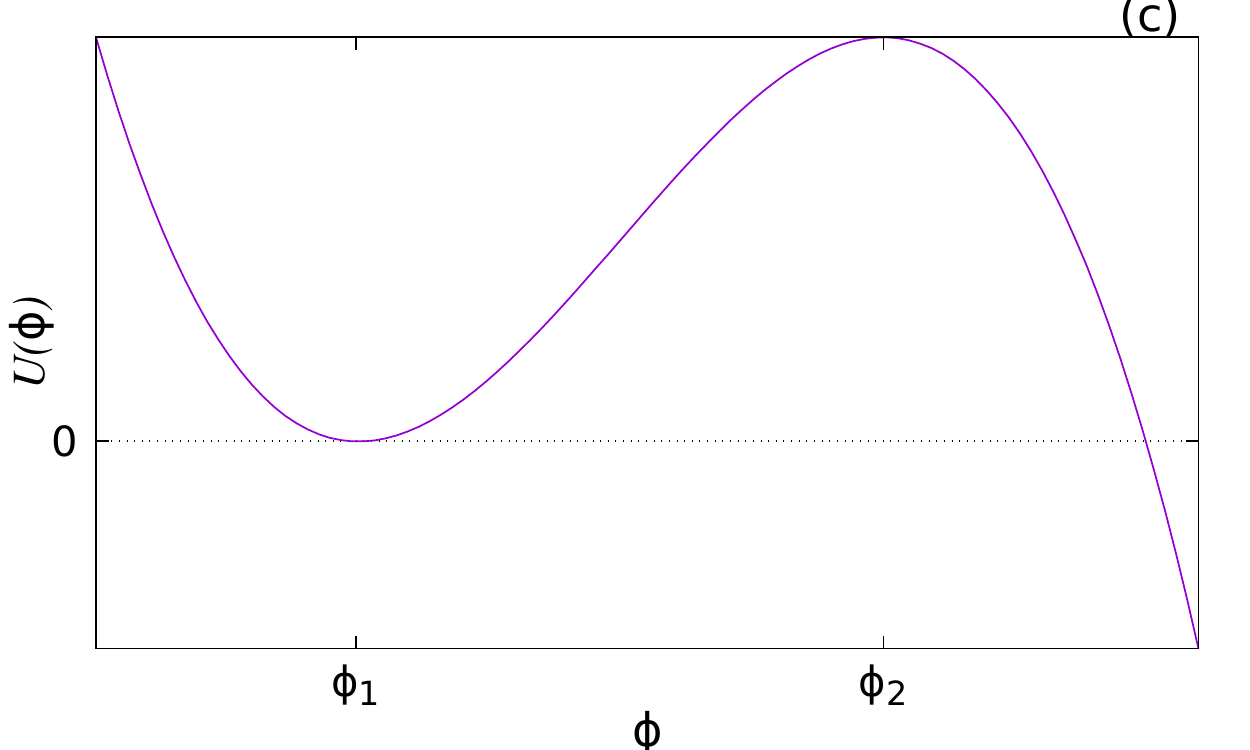}
\hfill
\includegraphics[width=.49\textwidth]{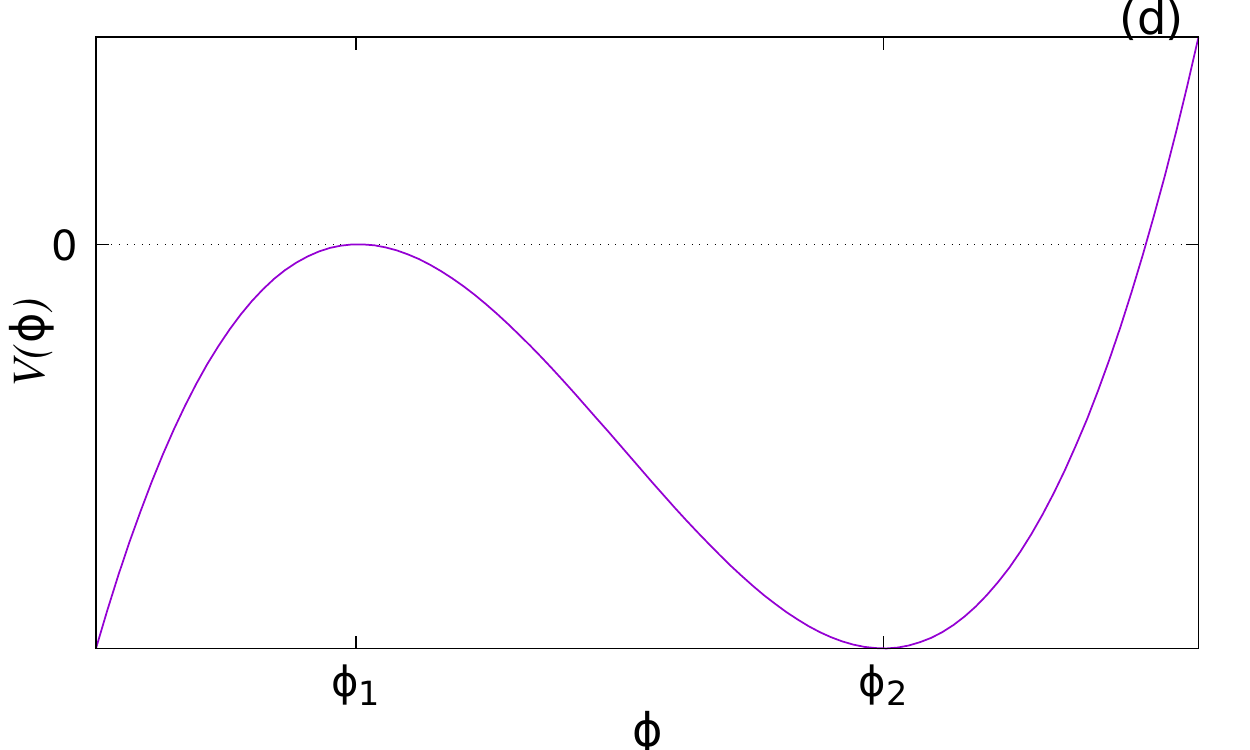}
\caption{\label{fig:1} \textbf{(a)} Potential $U(\phi)$ with a ``false'' vacuum and a ``true'' vacuum. \textbf{(b)} Potential
$V(\phi)=-U(\phi)$ for the motion of the fictitious particle. \textbf{(c)} Potential $U(\phi)$ showing a barrier separating a minimum
($\phi_1$) and a bottomless abyss. \textbf{(d)} Potential $V(\phi)=-U(\phi)$, where $U(\phi)$ has the shape shown in Fig.~\ref{fig:1}.(c).
}
\end{figure}

Recent experimental and theoretical works suggest that our vacuum is probably me\-tastable \cite{Grinstein2016, Sen2015, Burda2015, Ema2016,
Hoang2015, Degrassi2012, Burda2016, Butazzo2013, EliasMiro2012, Aad2012, Chatrchyan2012, Andreassen2014, Aad2015, Khachatryan2016, Bezrukov2012,
Aabound2016, Shkerin2015, Branchina2015, Khan2014, Ge2016, Kusenko2015, Hook2015, Bednyakov2015}. Measurements of Higgs boson and top
quark masses indicate that there should be a ``true'' vacuum with less energy than the present one \cite{Grinstein2016, Sen2015, Burda2015,
Ema2016, Hoang2015, Degrassi2012, Burda2016, Butazzo2013, EliasMiro2012, Aad2012, Chatrchyan2012, Andreassen2014, Aad2015,
Khachatryan2016, Bezrukov2012, Aabound2016, Shkerin2015, Branchina2015, Khan2014, Ge2016, Kusenko2015, Hook2015, Bednyakov2015, Kusenko2015b}. Vacuum stability problems have been discussed in many classical papers \cite{Coleman1977, Callan1977, Coleman1980,
Kobzarev1975, Gorsky2015, Blum2015, Isidori2001, Turner1982, Lindner1989, Sher1989, Krive1976, Cabibbo1979}. It is possible that
the current minimum of the scalar potential is local and a deeper minimum exists or the potential has a bottomless abyss
separated by a finite barrier [see Fig.~\ref{fig:1}]. In this situation, the Universe should eventually tunnel out into some
other state, in which the elementary particles and the laws of physics are different. According to current evidence, the Universe
is approximately 13.8 billion years old \cite{PlanckCol2016}. Scientists must find an explanation for these facts.
 
The framework for calculating the decay rate of a vacuum in Quantum Field Theory was developed by Coleman and collaborators
\cite{Coleman1977, Callan1977, Coleman1980}, inspired by the Kobzarev \textit{et al.} pioneering paper \cite{Kobzarev1975}.
For the vacuum to decay, it must tunnel through an energy barrier, and the probability, per unit time, per unit volume for
this has an exponential factor,
\begin{equation}
 \label{Eq01}
 \frac{\Gamma}{V}=Ae^{-\beta/\hslash},
\end{equation}
where $\beta$ is the action of a solution to the Euclidean field equations (the bounce) which interpolates between metastable (false) and true
vacua. This problem is related to many phenomena in condensed matter, particle physics and cosmology \cite{Cohen1993, Trodden1999, Yagi2005,
Linde1990, Weinberg2008, Gani2016, Hanggi1988, Aubry1975, Marchesoni1998}.

In this article we show that it is possible to have a stable false vacuum in a potential that is unbounded from below. The
remaining of the article is organized as follows. In Section \ref{model}, we briefly review the Coleman's model
\cite{Coleman1977}. In Section \ref{bounce}, we investigate the bounce solutions for some Lagrangians. In Section
\ref{discussion}, we discuss all the details of the actual evolution of the bubbles. We also present the results of numerical
simulations. Finally, in Section \ref{conclusions}, we present the conclusions of our analysis.

\section{The model}
\label{model}

Consider a Quantum Field Theory described by the Lagrangian
\begin{equation}
 \label{Eq02}
 \mathcal{L}=\frac{1}{2}\left(\partial_{\mu}\phi\right)\left(\partial^{\mu}\phi\right)-U(\phi),
\end{equation}
where $U(\phi)$ possesses two minima $\phi_1$, $\phi_3$ and a maximum $\phi_2$: $\phi_1<\phi_2<\phi_3$, $U(\phi_3)<U(\phi_1)$, $U(\phi_1)=0$ [see
Fig.~\ref{fig:1}.a]. The classical field equation would be
\begin{equation}
 \label{Eq03}
 \phi_{tt}-\nabla^2\phi=-\frac{dU(\phi)}{d\phi}.
\end{equation}

The state $\phi=\phi_1$ is a false vacuum. It is unstable by quantum effects (e.g. barrier penetration).
The accepted picture is the following. Quantum fluctuations are continually creating bubbles of true vacuum to materialize in the
false vacuum. Once in a while, a bubble of true vacuum will form large enough so that it is classically energetically favorable for
 the bubble to grow. The bubble expands throughout the Universe converting false vacuum to true.
 
 The Euclidean equation of motion is
 \begin{equation}
  \label{Eq04}
  \phi_{\tau\tau}+\nabla^2\phi=\frac{dU}{d\phi},
 \end{equation}
where $\tau=it$ is the imaginary time. The boundary conditions of the bounce solution are
\begin{eqnarray}
\label{Eq05}
  \phi_{\tau}(0, \mathbf{x})& = & 0,\\
\label{Eq06}
  \lim_{\tau\to\pm\infty}\phi(\tau,\mathbf{x}) & = &   \phi_1.
\end{eqnarray}
The integral that yields the action of the bounce is
\begin{equation}
 \label{Eq07}
 \beta=\int\,d\tau d^3x\left[\frac{1}{2}\phi_\tau^2+\frac{1}{2}\left(\boldsymbol{\nabla}\phi\right)^2+U\right].
\end{equation}
The action $\beta$ will be finite only if
\begin{equation}
\label{Eq08}
 \lim_{|\mathbf{x}|\to\infty}\phi(\tau,x)=\phi_1.
\end{equation}

The bounce solution is invariant under four-dimensional Euclidean rotations. Define
\begin{equation}
\label{Eq09}
 \rho\equiv(|\mathbf{x}|^2+\tau^2)^{1/2}.
\end{equation}
The bounce solution is a function of $\rho$ only. From equation \eqref{Eq04}, we obtain
\begin{equation}
\label{Eq10}
 \frac{d^2\phi}{d\rho^2}+\frac{3}{\rho}\frac{d\phi}{d\rho}=-\frac{dV(\phi)}{d\phi},
\end{equation}
where $V(\phi)=-U(\phi)$ [see Fig.~\ref{fig:1}.b]. The conditions for the bounce solutions are now
\begin{eqnarray}
\label{Eq11}
  \phi_{\rho}(0)& = & 0,\\
\label{Eq12}
  \lim_{\rho\to\infty}\phi(\rho) & = &   \phi_1.
\end{eqnarray}
The action $\beta$ can be calculated using the formula
\begin{equation}
 \label{Eq13}
 \beta=2\pi^2\int_0^{\infty}d\rho\,\rho^3\left[\frac{1}{2}\phi_{\rho}^2+U(\phi)\right].
\end{equation}

We can consider equation \eqref{Eq10} as the Newton's equation of motion for a fictitious particle of unit mass moving in the potential
$V(\phi)$ [see Fig.~\ref{fig:1}.b], where $\phi$ is the ``particle'' position and $\rho$ is the ``time''. There is a dissipative force acting on
the particle
\begin{equation}
 \label{Eq14}
 F_{dis}=-\frac{3}{\rho}\frac{d\phi}{d\rho}.
\end{equation}
The ``particle'' must be released at rest at time zero. The initial condition must be
\begin{equation}
 \label{Eq15}
 \phi_m<\phi_{(0)}<\phi_3.
 \end{equation}

For the bounce solution to exist, the ``particle'' should come to rest at ``time'' infinity at $\phi_1$ (i.e. on top of the maximum
$V(\phi_1)$) [see Fig.~\ref{fig:1}.b].

In many articles, the authors use the so-called thin-wall approximation \cite{Coleman1977} and the $\phi^4$ theory as
a model. The employed solution is nothing else but the one-dimensional $\phi^4$ soliton. This approach loses all the complexities
of the 3+1 dimensional nonlinear field equations. In the present paper we will show that, for some Lagrangians, the action of the
bounce solution can be arbitrarily large.

\section{The bounce solution}
\label{bounce}

Let us return to the motion of the fictitious particle in the potential $V(\phi)$ [see Fig.~\ref{fig:1}.b]. A particle that
starts its motion at a point $\phi(0)>\phi_m$ should pass the point $\phi=\phi_2$ (at the bottom of the valley) in order to reach the
point $\phi=\phi_1$. We can expect that in a vicinity of point $\phi_2$ the particle makes oscillations. This is true when the
potential $V(\phi)$ behaves as $V(\phi)\sim(\phi_2-\phi)^2/2$ in a neighborhood of point $\phi=\phi_2$, and we are studying the one-dimensional
case. For the $D=3+1$ case, the particle will make damped oscillations near the point $\phi=\phi_2$. Nevertheless, the
bounce solution can still exist.

Suppose there are potentials for which the motion of the ``particle'' in a vicinity of $\phi=\phi_2$ will be in the overdamped regime.
Then if the particle starts its motion at a point $\phi(0)>\phi_2$, there will be an overdamped trajectory that will end at point
$\phi=\phi_2$ for $\rho\to\infty$. The particle will never visit the points for which $\phi<\phi_2$.

Let us define
\begin{equation}
 \label{Eq16}
 U(\phi)=\left\{\begin{array}{lcc}
 \displaystyle\frac{a_1(\phi-\phi_1)^2}{2}, & \text{for} & \phi<\phi_{12},\\
 \displaystyle-\frac{(\phi-\phi_2)^4}{4}+\Delta_2, & \text{for} & \phi_{12}<\phi,
 \end{array}\right.
\end{equation}
where $\phi=\phi_{12}$ is a point for which $U(\phi)$ and its first derivative are continuous, i.e.
$a_1(\phi_{12}-\phi_1)^2/2 = \Delta_2-(\phi_{12}-\phi_2)^4/4$, and
$a_1(\phi_{12}-\phi_1)=-(\phi-\phi_2)^3$. All the solutions with the conditions
\begin{subequations}
\label{Eq17ab}
\begin{align}
\label{Eq16a}
\phi(0)>\phi_2,\\
\label{Eq16b}
\left.\frac{d\phi}{d\rho}\right|_{\rho=0}=0,
\end{align}
\end{subequations}
are given by the family of functions
\begin{equation}
 \label{Eq17}
 \phi(\rho)=\phi_2+\frac{Q}{1+\frac{Q^2}{8}\rho^2},
\end{equation}
where $\phi(0)=\phi_2+Q$. This includes solutions with the initial conditions
\begin{subequations}
\label{Eq19ab}
\begin{align}
\label{Eq18a}
 \phi(0)&>\phi_m,\\
 \label{Eq18b}
 \left.\frac{d\phi}{d\rho}\right|_\rho&=0.
\end{align}
\end{subequations}

Note that the solution \eqref{Eq17} is the most general solution satisfying the conditions \eqref{Eq18a} and \eqref{Eq18b}. If there
is another solution, there is no ``unicity''. Note also that there exists a continuum of solutions provided by \eqref{Eq17}, specified by any
initial condition $\phi(0)>\phi_2$.

It does not matter how high the ``particle'' will be at $\rho=0$ in the potential $V(\phi)$ [see Fig.~\ref{fig:1}.b], the
particle will be stuck at $\phi=\phi_2$ even for $\rho\to\infty$. It will never visit points close to $\phi=\phi_1$. The action of
all these solutions is infinite.

In general, for any potential that behaves as $U(\phi)\simeq(\phi-\phi_2)^4/4$ in a neighborhood of $\phi=\phi_2$, the motion of the
``particle'' for $D=3+1$ will be difficult near the point $\phi=\phi_2$.

Consider a generalization to the equation \eqref{Eq10} in a $D$-dimensional space-time
\begin{equation}
 \label{Eq20}
 \frac{d^2\phi}{d\rho^2}+\frac{D-1}{\rho}\frac{d\phi}{d\rho}=-\frac{d V(\phi)}{d\phi},
\end{equation}
where
\begin{equation}
\label{Eq21}
  U(\phi)=-V(\phi)=\frac{a|\phi|^3}{3}-\frac{b|\phi|^4}{4},\quad\quad a>0,\,b>0.
\end{equation}
For this potential, $\phi_1=0$, and $\phi_2=a/b$. We can find the exact bounce solution to equation \eqref{Eq20}
\begin{equation}
 \label{Eq22}
 \phi_{\text{bounce}}=\frac{Q}{1+N\rho^2},
\end{equation}
where
\begin{subequations}
\label{Eq23}
\begin{align}
\label{Eq23a}
 Q&=\frac{4}{4-D}\,\frac{a}{b},\\
\label{Eq23b}
 N&=\frac{2}{(4-D)^2}\,\frac{a^2}{b}.
 \end{align}
\end{subequations}
Notice that the bounce solution is unique in the sense that there is only one initial condition $\phi(0)=4a/(4-D)b$ for which
$\lim_{\rho\to\infty}\phi(\rho)=\phi_1\equiv0$.

We would like to remark that as $D$ is increased from $D=1$, the ``amplitude'' $Q$ of the bounce will increase (see eq.~\eqref{Eq23a})
($\phi(0)=Q$). In other words, we need an initial condition with enough potential energy for the ``particle''
 to be able to climb the hill and reach the point $\phi=\phi_1$. For larger $D$, there is more dissipation. So the particle needs more
 energy. However, $Q\to\infty$ as $D\to4$ (!). Something very important happens when $D=3+1$. This is an indication that for $D=4$, it is
 very difficult for the ``particle'' to arrive at point $\phi=\phi_1=0$. This can be also an evidence that
 \begin{equation}
  \label{Eq24}
  \lim_{D\to4}\beta_{\text{(bounce)}}=\infty.
 \end{equation}
This implies that there exists no finite bounce solution for $D=4$.

It is interesting to study Lagrangians where the potential $U(\phi)$ behaves as
\begin{subequations}
\label{Eq26}
\begin{align}
\label{Eq26a}
U(\phi)&\simeq\frac{(\phi-\phi_1)^{n_1}}{n_1}, \quad\text{near }\phi=\phi_1,\\
\label{Eq26b}
U(\phi)&\simeq\frac{(\phi_2-\phi)^{n_2}}{n_2}+\Delta_2, \quad\text{near }\phi=\phi_2,
\end{align}
and
\begin{equation}
 \label{Eq26c}
 U(\phi)\simeq\frac{(\phi-\phi_3)^{n_3}}{n_3}-\Delta_3, \quad\text{near }\phi=\phi_3,\\
\end{equation}
\end{subequations}
where $n_1\geq2$, $n_2\geq2$, and $n_3\geq2$. As $n_1$, $n_2$, $n_3$ and the dimension $D$ increase, the probability of the formation
of bubbles of the ``true'' vacuum inside the ``false'' vacuum that will expand approaches zero.
Compare the results of this Section with the results discussed in Ref.~\cite{Aravind2015}.

In brief, we gave for the first time, a mathematical proof that it is possible to have a stable false vacuum in a potential that is unbounded from below.

\section{Discussion}
\label{discussion}

Consider a chain of coupled pendula. In the continuum limit, this system can be described by the sine-Gordon equation
\begin{equation}
 \label{Eq27}
 \phi_{tt}-\phi_{xx}+\sin\phi=0.
\end{equation}

The static kink solution to equation \eqref{Eq27} is the following
\begin{equation}
\label{Eq28}
 \phi_k=4\arctan[\exp(x)].
\end{equation}
This solution is stable \cite{Gonzalez1998, Gonzalez2002, Gonzalez2007, Oliveira1996, GarciaNustes2012}. However, note that there
are pendula that are completely outside the stable equilibrium $\phi=0$ ($\phi=2\pi$). In fact, for the point $x=0$, the pendulum is at
position $\phi=\pi$, which, for a single pendulum, would be a completely unstable position. In the solution given by equation
\eqref{Eq28}, the pendulum at position $\phi=\pi$ and those that are close to that one are sustained by the majority of the pendula,
which are close to the stable position $\phi=0$ ($\phi=2\pi$).

Consider now a chain of linked small balls, which are moving in the potential
\begin{equation}
\label{Eq29}
U(\phi)=\frac{\phi^2}{2}-\frac{\phi^3}{3}.
\end{equation}
The graph of this function is similar to Fig.~\ref{fig:1}.c. The equation of motion is
\begin{equation}
 \label{Eq30}
 \phi_{tt}-\phi_{xx}=-\phi+\phi^2,
\end{equation}
whose exact critical-bubble solution \cite{Oliveira1996} can be obtained as
\begin{equation}
 \label{Eq31}
 \phi(x)=\frac{3}{2\cosh^2(x/2)}. 
\end{equation}

The unstable equilibrium of the potential is $\phi_2=1$. For instance, an initial condition of the form
\begin{subequations}
\label{Eq32}
\begin{align}
\label{Eq32a}
 \phi(x,0)&=\frac{1.4}{\cosh^2(x)},\\
 \label{Eq32b}
 \phi_t(x,0)&=0,
\end{align}
\end{subequations}
would lead to a dynamics where the balls fall to the ``left'' where the potential well ($\phi_1=0$) exists. There are balls situated to the
``right'' of the unstable equilibrium. However, these balls will not fall to the ``right'' (to the ``abyss''). The 
ball outside the potential well are sustained by the balls that are inside the potential well. We should say that ``larger'' bubbles
like the following
\begin{equation}
 \label{Eq33}
 \phi(x,0)=\frac{1.6}{\cosh^2(x/4)},
\end{equation}
will grow for ever.

Let us examine the following equation
\begin{equation}
\label{Eq34}
 \phi_{tt}-\nabla^2\phi=F(\phi),
\end{equation}
where $F(\phi)=a\left[\phi(\phi+2)(2+\delta-\phi)\right]^{2n-1}$, and $n\geq3$. The potential $U(\phi)$ ($F(\phi)=-dU(\phi)/d\phi$) for
equation \eqref{Eq34} has the shape shown in Fig.~\ref{fig:1}.a with two minima (a ``false'' vacuum and a ``true'' vacuum) and the properties
discussed in Section \ref{bounce}. In $D=3+1$ dimensions, the ``balls'' that are close
to the unstable equilibrium position $\phi_2=0$ are capable to sustain the ``balls'' that are very far away from the equilibrium
(including amplitudes that are arbitrarily large).
Initial conditions like the following
\begin{subequations}
\label{Eq35}
\begin{align}
\label{Eq35a}
 \phi(x,y,z,0)&=\phi_o+\frac{Q}{[1+b(x^2+y^2+z^2)]^{1/2}},\\
 \label{Eq35b}
 \phi_t(x, y, z,0)&=0,
\end{align}
\end{subequations}
with $Q>0$ and $b>0$, where all the ``balls'' are outside the unstable equilibrium, do not lead to an evolution with the ``balls''
falling to the ``right''.

Note that there are ``bubbles'', whose action is infinite and are made mostly of ``true'' vacuum, but, nevertheless, do
not grow. This means they do not expand throughout the Universe converting false vacuum to true. This leads to the conclusion that
``normal'' bubbles with finite action and made of ``true'' vacuum inside the Universe of ``false'' vacuum will not grow either.

Let us study an initial condition where the majority of the ``balls'' are close to the locally stable equilibrium $\phi=\phi_1$. In
$D=3+1$, the system is able to sustain any field configuration with the values of $\phi$ as far to the right as we wish from the
unstable equilibrium $\phi=\phi_2$. In $D=3+1$ the link between the ``balls'' is so strong that the ``balls'' outside the
equilibrium cannot drag those that are inside the equilibrium. Initial conditions in the form
\begin{subequations}
\label{Eq36}
\begin{align}
\label{Eq36a}
 \phi(x,y,z,0)&=-2+A\left[\tanh[B(r+d)]-\tanh[B(r-d)]\right],\\
 \label{Eq36b}
 \phi_t(x, y, z,0)&=0,
\end{align}
\end{subequations}
will not evolve in such a way that the bubble will expand, and the false vacuum will be converted to the ``true'' vacuum.

\begin{figure}[tbp]
\centering
\includegraphics[width=.49\textwidth]{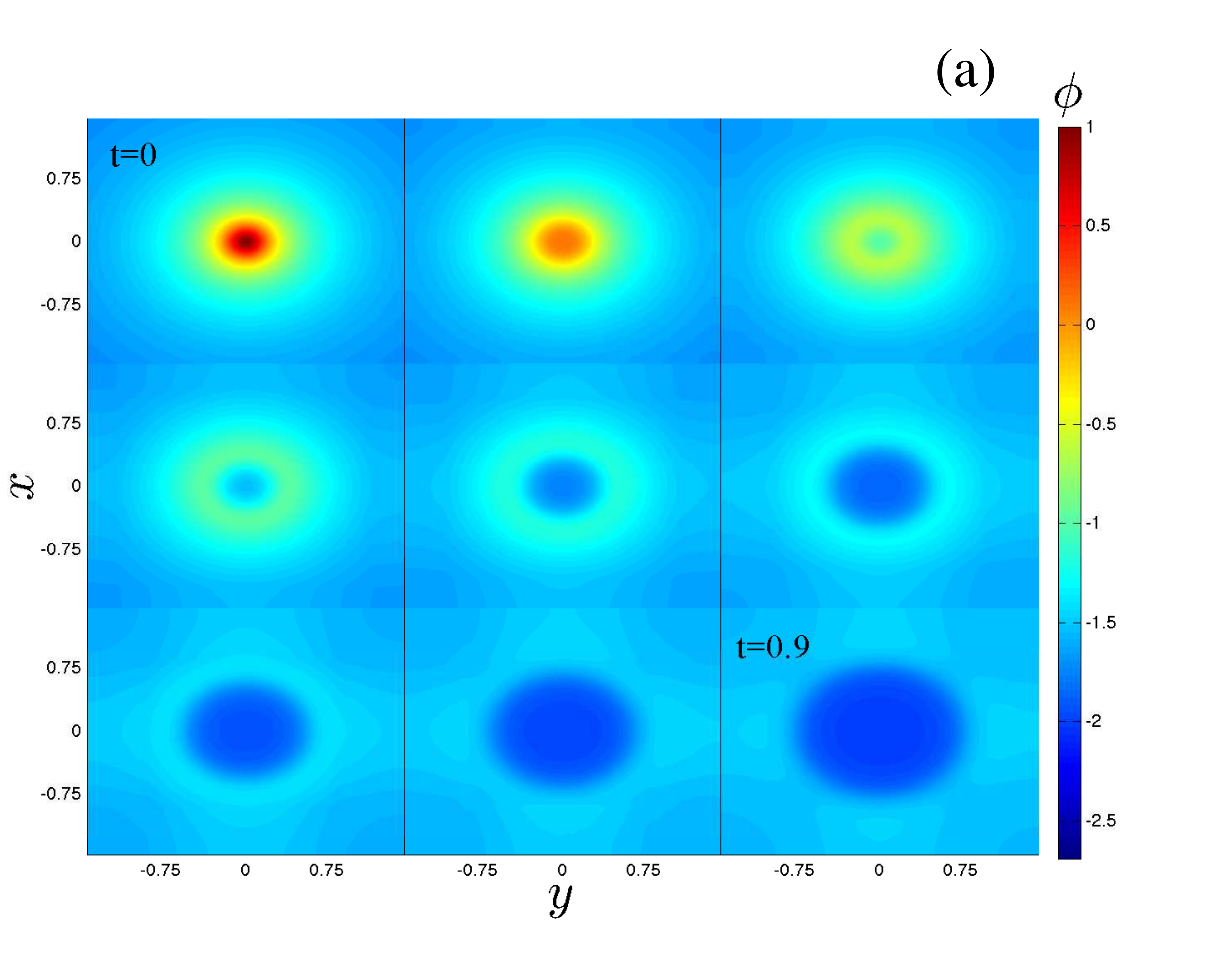}
\hfill
\includegraphics[width=.49\textwidth]{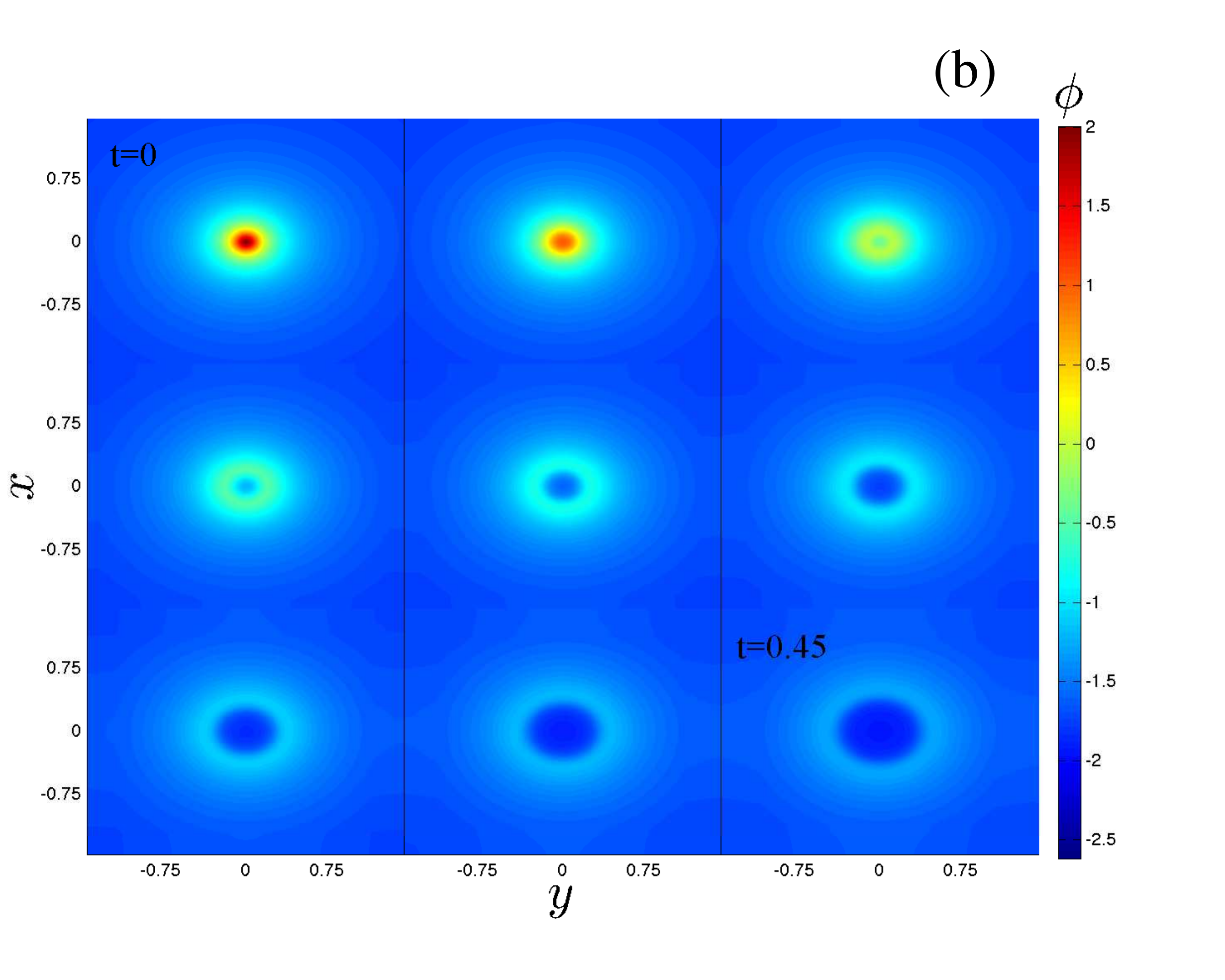}\\
\includegraphics[width=.49\textwidth]{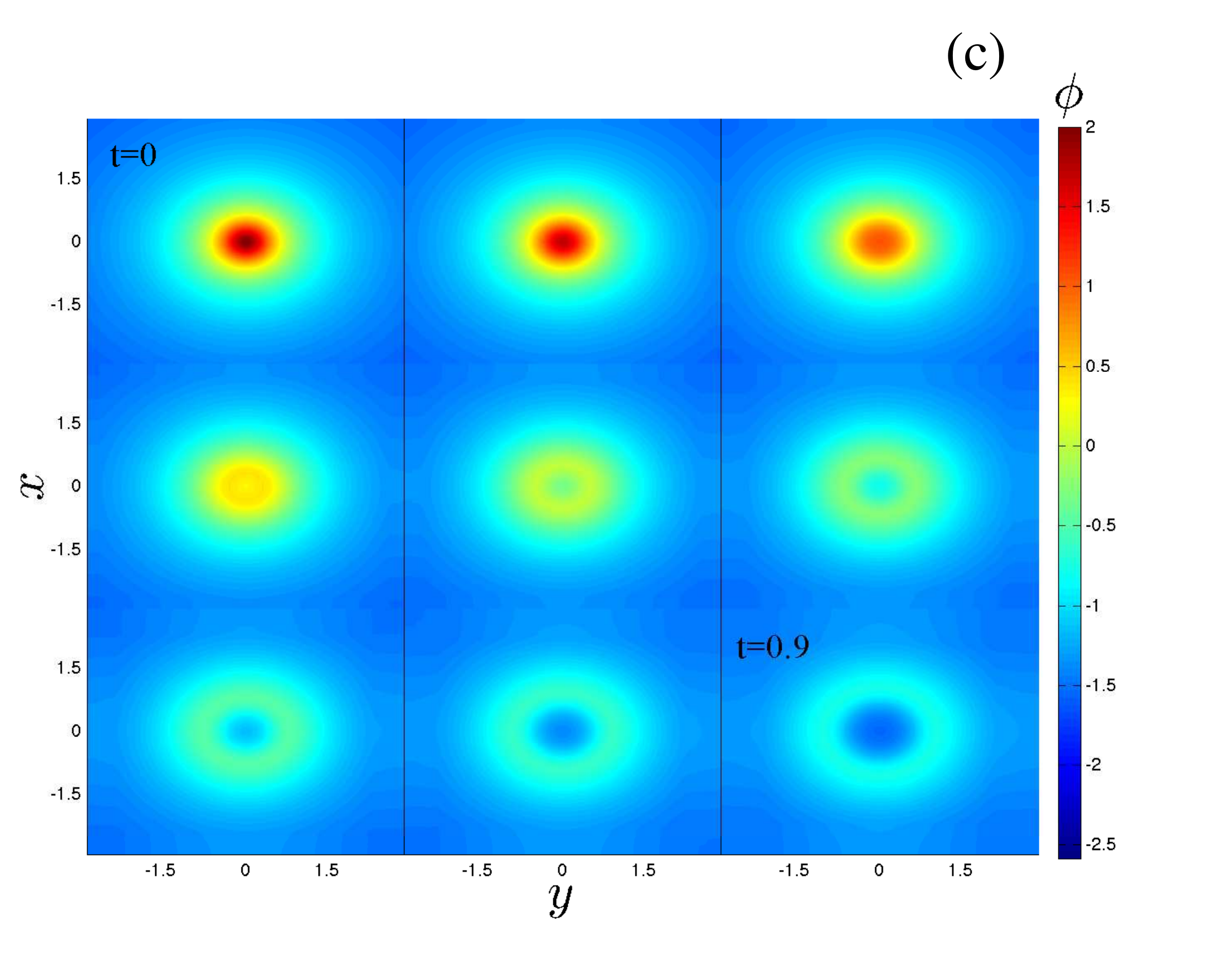}
\hfill
\includegraphics[width=.49\textwidth]{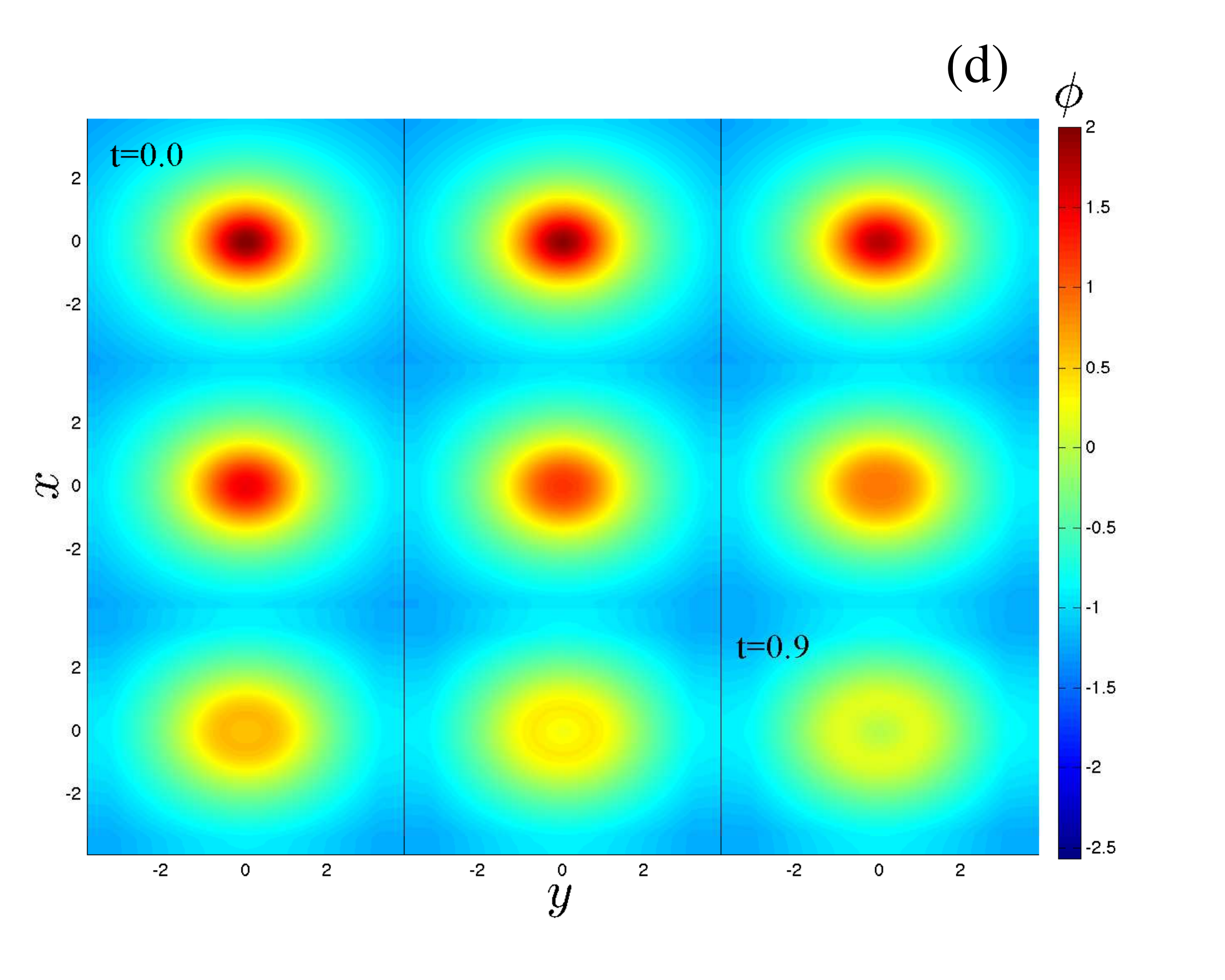}\\
\includegraphics[width=.49\textwidth]{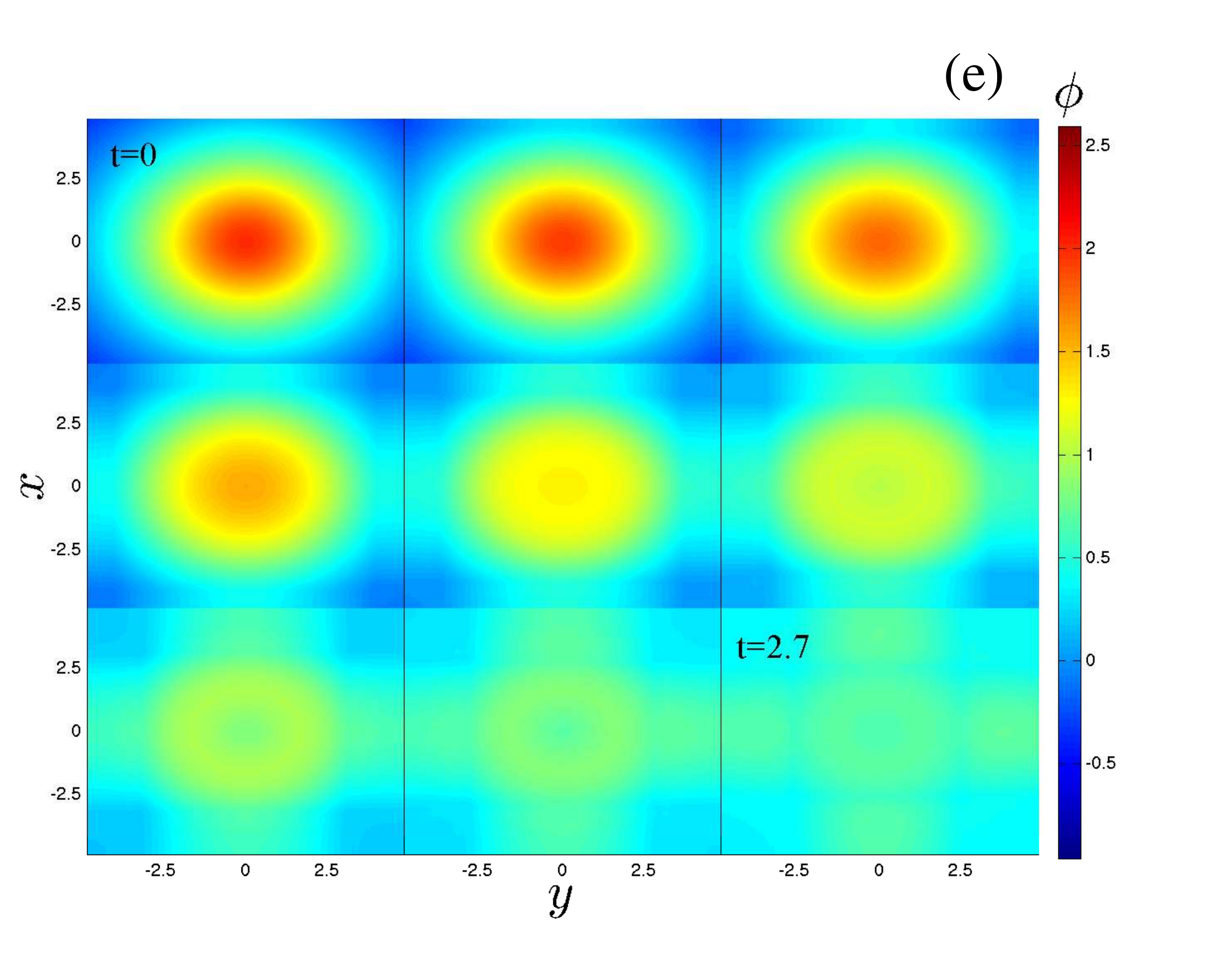}
\hfill
\includegraphics[width=.49\textwidth]{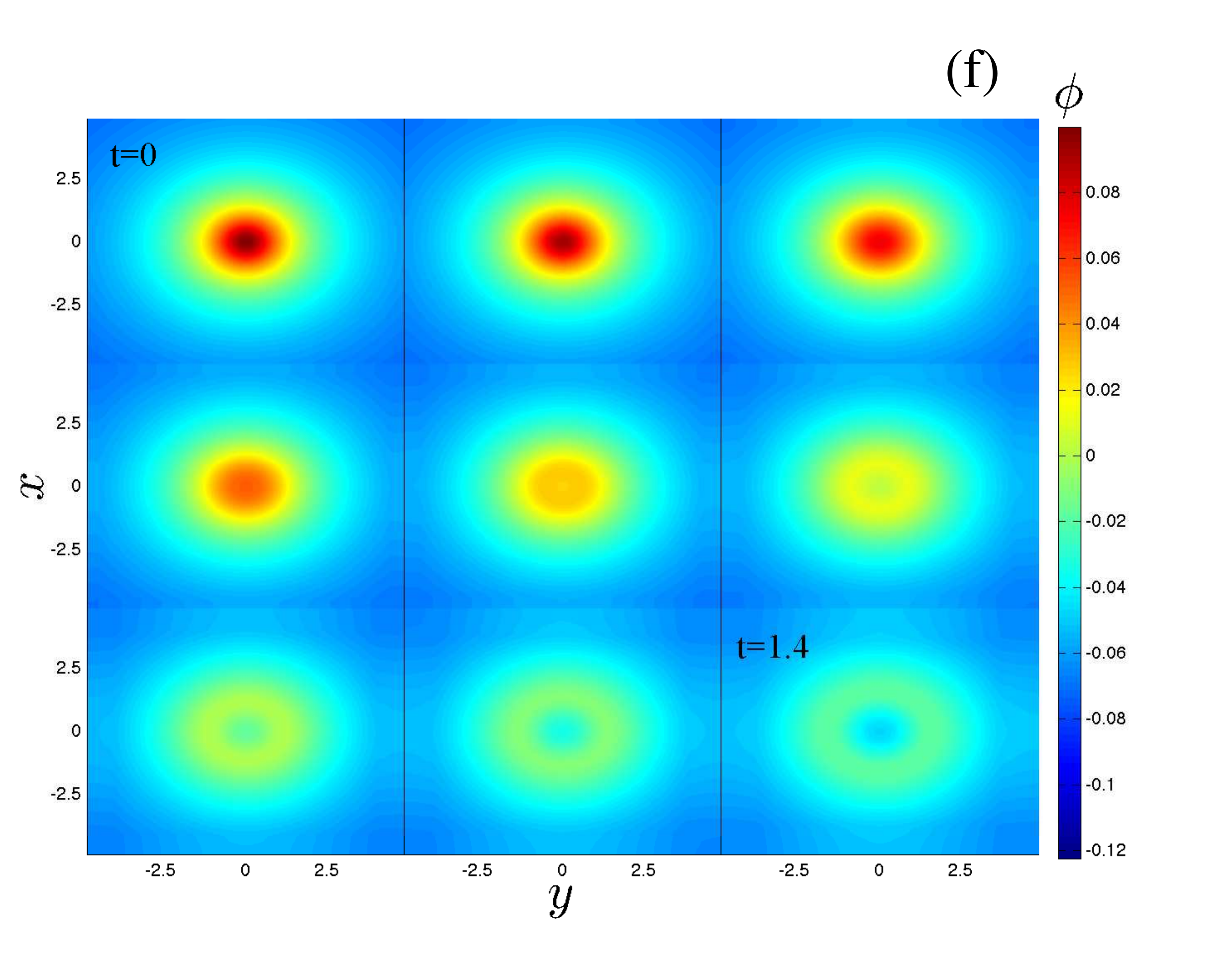}\\
\caption{\label{fig:2} Collapse of bubbles with different contents of ``true'' vacuum inside a Universe made of ``false'' vacuum. Results
obtained from the numerical solution of eq.~\eqref{Eq34} in 3 dimensions with the initial conditions of eq.~\eqref{Eq35} for \textbf{(a)} $\phi_o=-2$, $Q=3$,
and $b=27$. \textbf{(b)} $\phi_o=-2$, $Q=4$, and $b=85$. \textbf{(c)} $\phi_o=-2$, $Q=4.001$, and $b=5$. \textbf{(d)} $\phi_o=-2$, $Q=4.005$,
and $b=1$. \textbf{(e)} $\phi_o=-2$, $Q=4.001$, and $b=0.1$. \textbf{(f)} $\phi_o=-0.1$, $Q=0.2$, and $b=1$. Only the time evolution
of the field at $z=0$ is showed. Any other profile will exhibit a similar behavior due to the symmetry of the system.
}
\end{figure}

Figure \ref{fig:2} shows that most bubbles made of ``true'' vacuum inside a Universe made of ``false'' vacuum will collapse very fast.
There is a relaxation process such that $\lim_{t\to\infty}\phi=\phi_1$.

\begin{figure}[tbp]
\centering
\includegraphics[width=.49\textwidth]{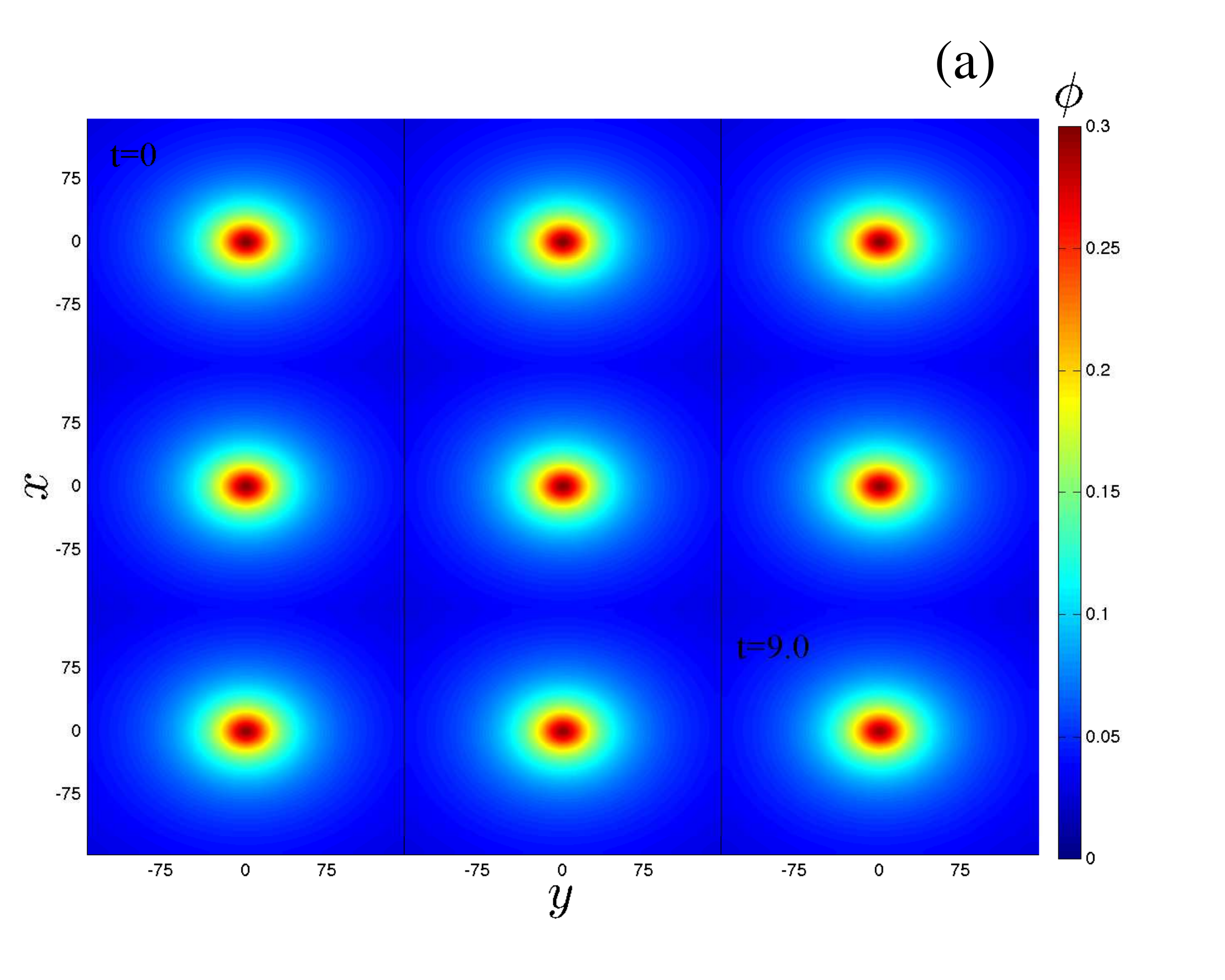}
\hfill
\includegraphics[width=.49\textwidth]{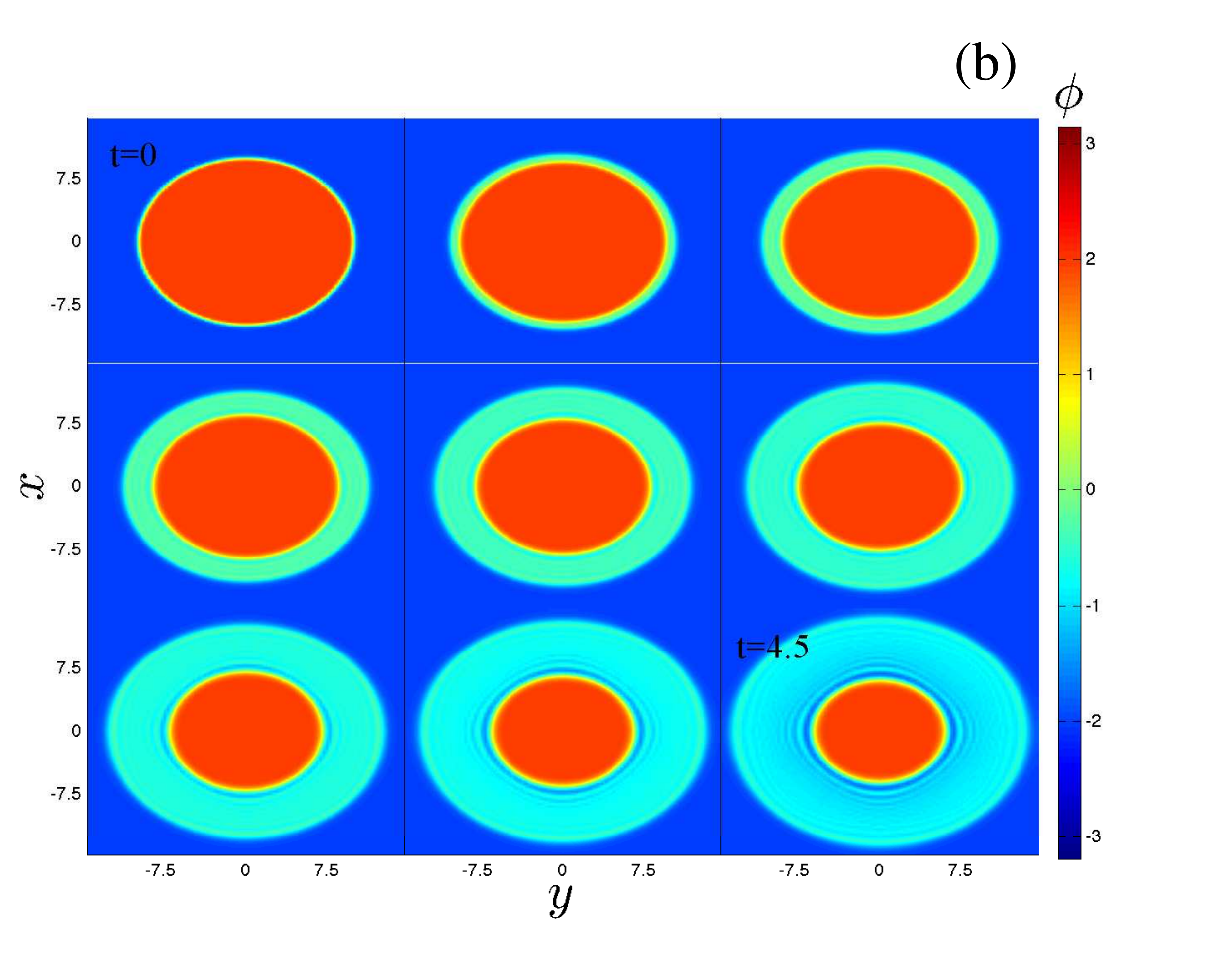}
\caption{\label{fig:3} \textbf{(a)} A stationary bubble in 3 dimensions mostly made of true vacuum. Results from numerical simulations of
eq.~\eqref{Eq34} for the initial conditions of eq.~\eqref{Eq35} for $\phi_o=0$, $Q=0.3$, and $b=0.0027$. \textbf{(b)} Results from numerical
simulations in 3 dimensions for the initial condition of Eq.~\eqref{Eq36} for $B=5$, $d=10$ and $A=2.0005$. Initially there is a
partial collapse of the walls of the field configuration, but there is no bubble of true vacuum that expands throughout the Universe. Only the time evolution
of the field at $z=0$ is showed. Any other profile will exhibit a similar behavior due to the symmetry of the system.
}
\end{figure}

Figure \ref{fig:3}.a shows the dynamics of a ``bubble'' made completely of ``true vacuum''. This bubble is stationary and marginally stable. There
is a whole continuum ``zone'' in configuration space with bubbles of different amplitudes and different behaviors of the tails. All the bubbles
in this ``zone'' are stationary solutions to the field equations. These bubbles are not asymptotically stable. A perturbation of the bubble will
lead to a different stationary configuration that belongs to the mentioned ``zone''. However, the most remarkable fact is that these bubbles do
not expand!

Figure \ref{fig:3}.b shows the dynamics of a very large bubble made of ``true'' vacuum inside a Universe of ``false'' vacuum. This bubble does
not expand. However, it is so large that its collapse is very slow. This bubble can be considered a quasi-stationary long-lived structure that
could contain ``new physics''. The properties and physical meaning of these structures will be discussed elsewhere.
All bubbles of the ``true'' vacuum inside a Universe of the ``false'' vacuum will collapse.

As a final remark, note that solutions \eqref{Eq17} are not bounce.
We have discussed them for some given potentials in Section \ref{bounce}.
However, we
have a system with an exact bounce solution (given by equations \eqref{Eq22} and \eqref{Eq23}) which is a unique solution.
There is only one initial condition such that for $\rho$ approaching
infinity, $\phi$ will be approaching $\phi_1$. The bounce solution exists
for $D<4$. However, as $D\to4$, $Q\to\infty$. This shows that $\lim_{D\to4}\beta_{bounce}=\infty$. Indeed, 
numerical investigations with the ordinary differential equation \eqref{Eq10} using the Strauss-Vazquez method \cite{Strauss1978, Jimenez2013}
show that the ``particle'' never reaches point $\phi=\phi_1$. Additionally, using the general partial differential equation \eqref{Eq34} we have
shown that the bubbles of true vacuum never expand. The potentials involved in these problems have degenerate critical points in the
sense of Ren\'e Thom \cite{Thom1975}, and the solutions near such points have special properties. This kind of potentials are relevant when we are
near criticality
\cite{Degrassi2012, Butazzo2013}.

\section{Conclusions}
\label{conclusions}

We have investigated the bounce solutions in some classes of vacuum decay problems. We have obtained a rigorous mathematical proof of the fact that
it is possible to have a stable false vacuum in a potential that is unbounded from below (See Section \ref{bounce}). This is our actual and main
result. 
Additionally, using the analysis of the results discussed in Sections \ref{bounce} and \ref{discussion}, and the numerical simulations, we can
formulate several conjectures. Note that the potential used in the numerical simulations (Eq.~\eqref{Eq34}) has the shape shown in Fig.~\ref{fig:1}.a
with two finite minima (a ``false'' vacuum and a ``true'' vacuum).

\begin{itemize}
 \item We conjecture that when the critical points of potential $U(\phi)$ are near criticality \cite{Degrassi2012, Butazzo2013} (see Sections
 \ref{bounce} and \ref{discussion}),
 the bounce solution does not exist or the action of the bounce solution is arbitrarily large. So the probability of the ``false'' vacuum decay is
 exponentially small.
 
 \item We conjecture that under the above conditions the bubbles of ``true'' vacuum inside a Universe of ``false'' vacuum will not grow. Thus there
 is no bubble expansion that would convert the ``false'' vacuum to ``true''.
 
 \item We conjecture (after Turner and Wilczek \cite{Turner1982}) that ``It is possible that our present vacuum is metastable, and that,
 nevertheless, the Universe would have chosen to get 'hung up' in it''.
\end{itemize}

\acknowledgments

The authors thanks M. Chacon for fruitful discussions and for his assistance with numerical simulations. J.F.M. thanks Saliya Coulibaly for his
most-valuable assistance with the semi-spectral methods used in part of the numerical simulations. M.A.G-N. thanks for the financial support of
Proyecto Puente PUCV. J.F.M. acknowledges financial
support of CONICYT doctorado nacional No. 21150292.



\begin{thebibliography}{99}




\bibitem{Grinstein2016} B. Grinstein and C. Murphy, \textit{Semiclassical approach to heterogeneous vacuum decay}, \textit{J. High Energy Phys.} \textbf{12},
063 (2015)
\bibitem{Sen2015} A. Sen, \textit{Riding gravity away from doomsday}, \textit{Int. J. Mod. Phys. D} \textbf{24}, 12:1544004 (2015)
\bibitem{Burda2015} P. Burda, R. Gregory and I. Moss, \textit{Gravity and the stability of the Higgs vacuum}, \textit{Phys. Rev. Lett.} \textbf{115},
  071303 (2015)
  \bibitem{Ema2016} Y. Ema, K. Mukaida and K. Nakayama, \textit{Fate of electroweak vacuum during preheating}, \textit{J. Cosmol. Astropart. Phys.} \textbf{10}, 043 (2016)
  \bibitem{Hoang2015} V. Hoang, P. Hung and A. Kamat, \textit{Non-sterile electroweak-scale right-handed neutrinos and the dual nature of the
  125-GeV scalar}, \textit{Nucl. Phys. B} \textbf{896}, 611 (2015) 
  \bibitem{Degrassi2012} G. Degrassi, S. Di Vita, J. Elias-Mir\'o, J. R. Espinosa, G. F. Giudice, G. Isidori and A. Strumia,
  \textit{Higgs mass and vacuum stability in the standard model at NNLO}, \textit{J. High Energy Phys.} \textbf{08}, 098 (2012) 
  \bibitem{Burda2016} P. Burda, R. Gregory and I. Moss, \textit{The fate of the Higgs vacuum}, \textit{J. High Energy Phys.} \textbf{06}, 025 (2016)
  \bibitem{Butazzo2013} D. Buttazzo, G. Degrassi, P. P. Giardino, G. F. Giudice, F. Sala, A. Salvio and A. Strumia, \textit{Investigating the
  near-criticality of the Higgs boson}, \textit{J. High Energy Phys.} \textbf{12}, 089 (2013)
  \bibitem{EliasMiro2012} J. Elias-Miro, J. R. Espinosa, G. F. Giudice, G. Isidori, A. Riotto and A. Strumia, \textit{Higgs mass implications on the stability of the electroweak vacuum}, \textit{Phys.
  Lett. B} \textbf{709}, 3:222 (2012)
  \bibitem{Aad2012} G. Aad \textit{et al.} (ATLAS Collaboration), \textit{Combined search for the standard model Higgs boson using up to
  $4.9$fb$^{-1}$ of $pp$ collision data at $\sqrt{s}=7TeV$ with the ATLAS detector at the LHC}, \textit{Phys. Lett. B} \textbf{710}, 1:49 (2012)
  \bibitem{Chatrchyan2012} S. Chatrchyan \textit{et al.} (CMS Collaboration), \textit{Combined results of searches for the standard model Higgs
  boson in pp collisions at $\sqrt{s}=7TeV$}, \textit{Phys. Lett. B} \textbf{710}, 1:26 (2012) 
  \bibitem{Andreassen2014} A. Andreassen, W. Frost and M. Schwartz, \textit{Consistent use of the standard model effective potential}, \textit{Phys. Rev.
  Lett.} \textbf{113}, 241801 (2014)
  \bibitem{Aad2015} G. Aad \textit{et al.} (ATLAS collaboration), \textit{Measurement of the top quark mass in the $t\bar{t}\to$lepton+jets and
  $t\bar{t}\to$dilepton channels using $\sqrt{s}=7TeV$ ATLAS data}, \textit{Eur. Phys. J. C} \textbf{75}, 330 (2015)
  \bibitem{Khachatryan2016} V. Khachatryan \textit{et al.} (CMS collaboration), \textit{Measurement of the top quark mass using proton-proton data at
  $\sqrt{(}s)=7$ and $8TeV$}, \textit{Phys. Rev. D} \textbf{93}, 072004 (2016)
  \bibitem{Bezrukov2012} F. Bezrukov, M. Yu. Kalmykov, B. A. Kniehl and M. Shaposhnikov, \textit{Higgs boson mass and new physics}, \textit{J. High Energy Phys.} \textbf{10}, 140 (2012)
  \bibitem{Aabound2016} Aabound \textit{et al.} (ATLAS collaboration), \textit{Measurement of the top quark mass in the $t\bar{t}\to$dilepton
  channel from $\sqrt{s}=8Tev$ ATLAS data}, \textit{Phys. Lett. B} \textbf{761}, 350 (2016)
  \bibitem{Shkerin2015} A. Shkerin and S. Sibiryakov, \textit{On stability of electroweak vacuum during inflation}, \textit{Phys. Lett. B}
  \textbf{746}, 257 (2015)
  \bibitem{Branchina2015} V. Branchina, E. Messina and M. Sher, \textit{Lifetime of the electroweak vacuum and sensitivity to Planck scale physics},
  \textit{Phys. Rev. D} \textbf{91}, 013003 (2015)
  \bibitem{Khan2014} N. Khan and S. Rakshit, \textit{Study of electroweak vacuum metastability with a singlet scalar dark matter}, \textit{Phys.
  Rev. D} \textbf{90}, 113008 (2014)
  \bibitem{Ge2016} S. Ge, H. He, J. Ren and Z. Xianyu, \textit{Realizing dark matter and Higgs inflation in light of LHC diphoton excess}, \textit{Phys. Lett.
  B} \textbf{757}, 480 (2016)
  \bibitem{Kusenko2015} A. Kusenko, L. Pearce and L. Yang, \textit{Postinflationary Higgs relaxation and the origin of matter-antimatter asymmetry},
  \textit{Phys. Rev. Lett.} 114, 061302 (2015)
  \bibitem{Hook2015} A. Hook, J. Kearney, B. Shakya and K. M. Zurek, \textit{Probable of improbable Universe? Correlating electroweak vacuum instability with the scale of
  inflation}, \textit{J. High Energy Phys.} \textbf{01}, 061 (2015)
  \bibitem{Bednyakov2015} A. V. Bednyakov, B. A. Kniehl, A. F. Pikelner and O. L. Veretin, \textit{Stability of the electroweak vacuum: Gauge independence and advanced precision},
  \textit{Phys. Rev. Lett.} \textbf{115}, 201802 (2015)
  \bibitem{Kusenko2015b} A. Kusenko, \textit{Are we on the brink of the Higgs abyss?}, \textit{Physics} \textbf{8}, 108 (2015)
  \bibitem{Coleman1977} S. Coleman, \textit{Fate of the false vacuum: Semiclassical theory}, \textit{Phys. Rev. D} \textbf{15}, 2929 (1977)
  \bibitem{Callan1977} 	C. Callan and S. Coleman, \textit{Fate of the false vacuum. II. First quantum corrections}, \textit{Phys. Rev. D}
  \textbf{16}, 1762 (1977)
  \bibitem{Coleman1980} S. Coleman and F. de Luccia, \textit{Gravitational effects on and of vacuum decay}, \textit{Phys. Rev. D} \textbf{21},
  12:3305 (1980)
  \bibitem{Kobzarev1975} I. Kobzarev, L. Okun and M. Voloshin, \textit{Bubbles in metastable vacuum}, \textit{Sov. J. Nucl. Phys.} \textbf{20}, 644
  (1975) [\textit{Yad. Fiz.} \textbf{20}, 1229 (1974)]
  \bibitem{Gorsky2015} A. Gorsky \textit{et al.}, \textit{Is the standard model saved asymptotically by conformal symmetry?}, \textit{J. Exp.
  Theor. Phys.} \textbf{120}, 3:399 (2015)
  \bibitem{Blum2015} K. Blum, R. D'Agnolo and J. Fan, \textit{Vacuum stability bounds on Higgs coupling deviations in the absence of new bosons},
  \textit{J. High Energy Phys.} \textbf{03}, 166 (2015)
  \bibitem{Isidori2001} G. Isidori, G. Ridolfi and A. Strumia, \textit{On the metastability of the standard model vacuum}, \textit{Nucl. Phys. B}
  \textbf{609}, 3:387 (2001)
  \bibitem{Turner1982} M. Turner and F. Wilczek, \textit{Is our vacuum metastable?}, \textit{Nature} \textit{298} 633 (1982)
  \bibitem{Lindner1989} M. Lindner, M. Sher and H. Zaglauer, \textit{Probing vacuum stability bounds at the Fermilab collider}, \textit{Phys. Lett.
  B} \textbf{228}, 139 (1989)
  \bibitem{Sher1989} M. Sher, \textit{Electroweak Higgs potential and vacuum stability}, \textit{Phys. Rep.} \textbf{179}, 5-6:273 (1989)
  \bibitem{Krive1976} I. Krive and A. Linde, \textit{On the vacuum stability in the $\sigma$ model}, \textit{Nucl. Phys. B} \textbf{117}, 1:265
  (1976)
  \bibitem{Cabibbo1979} N. Cabibbo, L. Maiani, G. Parisi and R. Petronzio, \textit{Bounds on the fermions and Higgs boson masses in grand unified theories}, \textit{Nucl.
  Phys. B} \textbf{158}, 2-3:295 (1979)
  
  \bibitem{PlanckCol2016} P. A. R. Ade \textit{et al.} (Planck Collaboration), \textit{Planck 2015 results - XIII. Cosmological
  parameters}, \textit{Astronomy \& Astrophysics} \textbf{594}, A13 (2016)
    
  \bibitem{Cohen1993} A. Cohen, B. Kaplan and A. Nelson, \textit{Progress in electroweak baryogenesis}, \textit{Annu. Rev. Nucl. Part. Sci.}
  \textbf{43}, 27 (1993)
  \bibitem{Trodden1999} M. Trodden, \textit{Electroweak baryogenesis}, \textit{Rev. Mod. Phys.} \textbf{71}, 1463 (1999)
  \bibitem{Yagi2005} K. Yagi, T. Hatsuda and Y. Miake, \textit{Quark-Gluon Plasma}, Cambridge University Press, Cambridge, UK (2005)
  \bibitem{Linde1990} A. Linde, \textit{Particle Physics and Inflationary Cosmology}, Oxford University Press, Oxford (2008)
  \bibitem{Weinberg2008} S. Weinberg, \textit{Cosmology}, Oxford University Press, Oxford (2008)
  \bibitem{Gani2016} V. Gani, M. Lizunova and R. Radomskiy, \textit{Scalar triplet on a domain wall: an exact solution}, \textit{J. High
  Energy Phys.} \textbf{04}, 043 (2016)
  \bibitem{Hanggi1988} P. H\"anggi, F. Marchesoni and P. Sodano, \textit{Nucleation of thermal sine-Gordon solitons: Effect of many-body
  interactions}, \textit{Phys. Rev. Lett.} \textbf{60}, 2563 (1988)
  \bibitem{Aubry1975} J. Aubry, \textit{A unified approach to the interpretation of displacive and order-disorder systems. I. Thermodynamical
  aspect}, \textit{J. Chem. Phys.} \textbf{62}, 3217 (1975)
  \bibitem{Marchesoni1998} F. Marchesoni, C. Cattuto and G. Costantini, \textit{Elastic strings in solids: Thermal nucleation}, \textit{Phys. Rev. B}
  \textbf{57}, 7930 (1998)
  
  \bibitem{Aravind2015} A. Aravind, B. S. DiNunno, D. Lorshbough and S. Paban, \textit{Analyzing multifield tunneling with exact bounce solutions},
 \textit{Phys. Rev. D} \textbf{91}, 2:025026 (2015)

  \bibitem{Gonzalez1998} J. A. Gonz\'alez, B. A. Mello, L. I. Reyes and L. E. Guerrero, \textit{Resonance phenomena of a solitonlike extended object in a bistable potential},
  \textit{Phys. Rev. Lett.} \textbf{80}, 1361 (1998)
  \bibitem{Gonzalez2002} J. A. Gonz\'alez, A. Bellor\'in and L. E. Guerrero, \textit{Internal modes of sine-Gordon solitons in the presence of
  spatiotemporal perturbations}, \textit{Phys. Rev. E} \textbf{65}, 065601(R) (2002)
  \bibitem{Gonzalez2007} J. A. Gonz\'alez, S. Cuenda and A. S\'anchez, \textit{Kink dynamics in spatially inhomogeneous media: The role of internal
  modes}, \textit{Phys. Rev. E} \textbf{75}, 036611 (2007)
  \bibitem{Oliveira1996} F. Oliveira and J. A. Gonz\'alez, \textit{Bond-stability criterion in chain dynamics}, \textit{Phys. Rev. B} \textbf{54}, 3954 (1996)
   \bibitem{GarciaNustes2012} M. Garcia-\~Nustes and J. A. Gonz\'alez, \textit{Formation of a two-kink soliton pair in perturbed sine-Gordon models
   due to kink-internal-mode instabilities}, \textit{Phys. Rev. E} \textbf{86}, 066602 (2012)
  %
  \bibitem{Strauss1978} W. Strauss and L. V\'azquez, \textit{Numerical solution of a nonlinear Klein-Gordon equation}, \textit{J. Comput. Phys.}
  \textbf{28}, 2:271--278 (1978)
  %
  \bibitem{Jimenez2013} S. Jim\'enez, J. A. Gonz\'alez and L. V\'azquez, \textit{Fractional Duffing's Equation and Geometrical Resonance},
  \textit{Int. J. Bifurcation Chaos} \textbf{23}, 05:1350089 (2013)
  %
  \bibitem{Thom1975} R. Thom, \textit{Structural Stability and Morphogenesis}, W. A. Benjamin INC., Reading, Massachusetts (1975)




\end{thebibliography}
\end{document}